\documentclass[11pt]{article}
\usepackage{ae}
\usepackage{graphicx}
\usepackage{amsmath}
\usepackage{amssymb}
\usepackage{amsthm}
\usepackage{amsfonts}
\usepackage{latexsym}
\usepackage{verbatim}
\usepackage{verbatim}
\usepackage{url}
\usepackage[normalem]{ulem}
\newtheorem{thrm}{Theorem}[section]

\newtheorem{lmm}[thrm]{Lemma}

\newtheorem{prpstn}[thrm]{Proposition}

\theoremstyle{definition}

\newtheorem{rmrk}[thrm]{Remark}
\newtheorem{xmpl}[thrm]{Example}

\theoremstyle{remark}
\newtheorem*{note}{Note}

\newcommand{\cA}{{\mathcal A}}

\newcommand{\ZZ}{{\mathbb Z}}

\newcommand{\NN}{{\mathbb N}}

\newcommand{\cAstar}{\mathcal{A}^*}
\newcommand{\cAplus}{\mathcal{A}^+}
\newcommand{\empt}{\varepsilon}
\newcommand{\cAw}{\mathcal{A}^{\omega}}

\newcommand{\rev}{\widetilde}

\newcommand{\bb}{\mathbf{b}}

\newcommand{\bw}{\mathbf{w}}
\newcommand{\bx}{\mathbf{x}}
\newcommand{\by}{\mathbf{y}}

\newcommand{\bs}{\mathbf{s}}
\newcommand{\bt}{\mathbf{t}}

\newcommand{\Ult}{\mbox{Ult}}
\newcommand{\eUlt}{\mbox{\emph{Ult}}}

\def\Alph{\textrm{Alph}}

\newcommand{\br}{\mathbf{r}}
\newcommand{\bl}{\mathbf{l}}

\newenvironment{proof2}[1]{\par\noindent {\it Proof of #1.} \rm}{ \
~~~$\fbox{}$ }

\numberwithin{equation}{section}

\author{Amy Glen\footnotemark[2] \and 
Florence Lev\'e\footnotemark[3] \and Gw\'ena\"el Richomme\footnotemark[3]
}
\title{Directive words of episturmian words: equivalences and normalization}

\date{February 26, 2008}

\begin{document}


\normalsize

\maketitle 

\footnotetext[2]{
\small Amy Glen  \\
\small LaCIM, Universit\'e du Qu\'ebec \`a Montr\'eal, \\
\small C.P. 8888, succursale Centre-ville,
Montr\'eal, Qu\'ebec, CANADA, H3C 3P8 \\
E-mail:
\texttt{amy.glen@gmail.com} (with the support of CRM-ISM-LaCIM)}
\footnotetext[3]{
\small F. Levé, G. Richomme\\
\small Universit\'e de Picardie Jules Verne,\\
\small Laboratoire MIS (Modélisation, Information, Systèmes)\\
\small 33, Rue Saint Leu, F-80039 Amiens cedex 1, FRANCE\\
E-mail: \texttt{\{florence.leve, gwenael.richomme\}@u-picardie.fr}
}

\hrule
\begin{abstract}
Episturmian morphisms constitute a powerful tool to study episturmian words. Indeed, any episturmian word can be infinitely decomposed over the set of pure episturmian morphisms. Thus, an episturmian word can be defined by one of its morphic decompositions or, equivalently, by a certain directive word. Here we characterize pairs of words directing a common episturmian word. We also propose a way to uniquely define any episturmian word through a normalization of its directive words. As a consequence of these results, we characterize episturmian words having a unique directive word.

\medskip

\noindent {\bf Keywords}: episturmian word; Sturmian word; Arnoux-Rauzy sequence; episturmian morphism; directive word.
\vspace{0.1cm} \\
MSC (2000): 68R15.
\end{abstract}
\hrule

\section{Introduction}

Since the seminal works of Morse and Hedlund \cite{MH1940},  {\em Sturmian words} have been widely studied and their beautiful properties are related to many fields like Number Theory, Geometry, Dynamical Systems, and Combinatorics on Words (see \cite{AS2003,mL02alge,nP02subs,jB07stur} for recent surveys). These infinite words, which are defined on a binary alphabet, have numerous equivalent definitions and characterizations. Nowadays most works deal with generalizations of Sturmian words to arbitrary finite alphabets. Two very interesting generalizations are very close: the {\em Arnoux-Rauzy sequences} (e.g., see \cite{pAgR91repr, jJgP02onac, nP02subs, rRlZ00agen}) and {\em episturmian words} (e.g., see \cite{xDjJgP01epis, jJgP02epis, jJgP04epis}). The first of these two families is a particular subclass of the second one. More precisely, the family of episturmian words is composed of  the Arnoux-Rauzy sequences, images of the Arnoux-Rauzy sequences by {\em episturmian morphisms}, and certain periodic infinite words. In the binary case, Arnoux-Rauzy sequences are exactly the Sturmian words whereas episturmian words include all recurrent {\em balanced} words, that is, periodic balanced words and Sturmian words (see \cite{aGjJgP06char,PV2006,gR07aloc} for recent results relating episturmian words to the balanced property). See also \cite{aGjJ07epis} for a recent survey on episturmian theory.

Episturmian morphisms play a central role in the study of these words. Introduced first as a generalization of Sturmian morphisms, Justin and Pirillo \cite{jJgP02epis} showed that they are exactly the morphisms that preserve the aperiodic episturmian words. They also proved that any episturmian word is the image of another episturmian word by some so-called {\em pure episturmian morphism}. Even more, any episturmian word can be infinitely decomposed over the set of pure episturmian morphisms.  This last property allows  an episturmian word to be defined by one of its morphic decompositions or, equivalently, by a certain {\em directive word}, which is an infinite sequence of rules  for decomposing the given episturmian word by morphisms. In consequence, many properties of episturmian words can be deduced from properties of episturmian morphisms. This approach is used for instance in \cite{vBcHlZ06init, aG06acha, fLgR04quas,gR07conj,gR07aloc,rRlZ00agen} and of course in the papers of Justin {\em et al.}
In Section~\ref{Ss:episturmian}, we recall useful results on episturmian words and explain the vision of morphic decompositions and directive words introduced by Justin and Pirillo in \cite{jJgP02epis}. 

An episturmian word can have several directive words.
The question: ``When do two words direct a common episturmian word?'' was considered in \cite{jJgP04epis}. Using a block-equivalence notion for directive words, Justin and Pirillo provided several results to answer this question in most cases (see Section~\ref{S:directiveWords}). In Section~\ref{S:equivalence}, we state a complete result characterizing the form of words directing a common episturmian word, without using block-equivalence.

In \cite{vBcHlZ06init}, Berth\'e, Holton, and Zamboni show that any Sturmian word has a unique directive word with some particular properties. In \cite{fLgR07quas}, the second and third authors rephrased this result and used it to characterize all quasiperiodic Sturmian words. In Section~\ref{S:normalization}, we extend this result to all episturmian words by introducing a way  to normalize the directive words of an episturmian word so that any episturmian word can be defined uniquely by its {\em normalized directive word}, defined by some factor avoidance (Theorem~\ref{T:normalisation}).  This result was previously presented at the {Sixth International Conference on Words} \cite{fLgR07quasB} to characterize all quasiperiodic episturmian words (see also \cite{aGfLgR07quas}).

As an application of the previous results, we end this paper with a characterization of episturmian words having a unique directive word.

\section{Episturmian words and morphisms\label{Ss:episturmian}}

We assume the reader is familiar with combinatorics on words and morphisms (e.g., see \cite{mL83comb,mL02alge}).  
In this section, we recall some basic definitions and properties relating to episturmian words which are  needed later in the paper.  For the most part, we follow the notation and terminology of \cite{xDjJgP01epis, jJgP02epis, jJgP04epis, aGjJgP06char}.

\subsection{Notation and terminology}

Let $\cA$ denote a finite {\em alphabet}. A finite \emph{word} over $\cA$ is a finite sequence of letters from $\cA$. The {\em empty word} $\empt$ is the empty sequence. Under the operation of concatenation, the set $\cA^*$ of all finite words over $\cA$ is a {\em free monoid} with identity element $\empt$ and set of generators $\cA$. The set of {\em non-empty} words over $\cA$ is the {\em free semigroup} $\cA^+ = \cAstar \setminus \{\empt\}$.  

Given a finite word $w = x_{1}x_{2}\cdots x_{m} \in \cAplus$ with each $x_{i} \in \cA$, the \emph{length} of $w$ is $|w| = m$. The length of the empty word is $0$. 
By $|w|_a$ we denote the number of occurrences of the letter $a$ in the word $w$. If $|w|_a = 0$, then $w$ is said to be {\em $a$-free}. For any integer $p \geq 1$, the $p$-th power of $w$ is the word $w^p$ obtained by concatenating $p$ occurrences of $w$.

A (right) \emph{infinite word} $\bx$ is a sequence indexed by $\NN^+$ with values in $\cA$, i.e., $\bx = x_1x_2x_3\cdots$ with each $x_i \in \cA$. The set of all infinite words over $\cA$ is denoted by $\cAw$. 
Given a non-empty finite word $v$, we denote by $v^\omega$ the infinite word obtained by concatenating $v$ with itself infinitely many times. For easier reading, infinite words are hereafter typed in boldface to distinguish them from finite words.

Given a set $X$ of words, $X^*$ (resp.~$X^\omega$) is the set
of all finite (resp.~infinite) words that can be obtained by 
concatenating words of $X$. The empty word $\varepsilon$ belongs to $X^*$.

A finite word $w$ is a \emph{factor} of a finite or infinite word $z$ if $z = uwv$ for some words $u$, $v$ (where $v$ is infinite iff $z$ is infinite). Further, $w$ is called a \emph{prefix} (resp.~\emph{suffix}) of $z$ if $u = \empt$ (resp.~$v = \empt$).  We use the notation $p^{-1}w$ (resp.~$ws^{-1}$) to indicate the removal of a prefix $p$ (resp.~suffix $s$) of the word~$w$.

The \emph{alphabet} of a word $w$, denoted by Alph$(w)$ is the set of letters occurring in $w$, and if $w$ is infinite, we denote by Ult$(w)$ the set of
all letters occurring infinitely often in $w$.

\subsection{\label{Ss:defEpisturmian}Episturmian words}

In this paper, our vision of episturmian words will be the characteristic property stated in Theorem~\ref{T:episturmian} below. Nevertheless, to give an idea of what an episturmian word is, let us give one of the equivalent definitions of an episturmian word provided in \cite{xDjJgP01epis}. 
Before doing so, we recall that a factor $u$ of an infinite word $\bw \in \cA^\omega$ is \emph{right} (resp.~\emph{left}) \emph{special} if $ua$, $ub$ (resp.~$au$, $bu$) are factors of $\bw$ for some letters $a$, $b \in \cA$, $a \ne b$. We recall also that the \emph{reversal} $\rev{w}$ of a finite word $w$ is its mirror image: if $w = x_1\ldots x_{m-1}x_m$, then $\rev{w} = x_{m}x_{m-1}\cdots x_{1}$.

An infinite word $\bt \in \cAw$ is \emph{episturmian} if its set of factors is closed under reversal and $\bt$ has at most one right (or equivalently left) special factor of each length. Moreover, an episturmian word is \emph{standard} if all of its left special factors are prefixes of it.

In the initiating paper \cite{xDjJgP01epis}, episturmian words were defined in two steps. Standard episturmian words were first introduced and studied as a generalization of standard Sturmian words.  (Note that in the rest of this paper, we refer to a standard episturmian word as an {\em epistandard word}, for simplicity). 
Then an episturmian word was defined as an infinite word having exactly the same set of factors as some epistandard word. 

Moreover, it was proved in \cite{xDjJgP01epis} that any episturmian word is \emph{recurrent}, that is, all of its factors occur infinitely often (actually episturmian words are uniformly recurrent but this will not be needed here).
An {\em ultimately periodic} infinite word is a word that can be written as $uv^\omega = uvvv\cdots$, for some $u$, $v \in \cAstar$, $v\ne \empt$. If $u = \empt$, then such a word is {\em periodic}. Since they are recurrent, all ultimately periodic episturmian words are periodic.
 Let us recall that an infinite word that is not ultimately periodic is said to be {\em aperiodic}.

\subsection{Episturmian morphisms} \label{SS:EpiMorphisms}

To study episturmian words, Justin and Pirillo \cite{jJgP02epis} introduced {\em episturmian morphisms}. 
In particular they proved that these morphisms (defined below) are precisely the morphisms that preserve the set of aperiodic episturmian words.

\medskip

Let us recall that given an alphabet $\cA$, a \textit{morphism} $f$ on $\cA$ is a map from $\cA^*$ to $\cA^*$ such that $f(uv) = f(u) f(v)$ for any words $u$, $v$ over $\cA$.  A morphism on $\cA$ is entirely defined by the images of letters in $\cA$. All morphisms considered in this paper will be non-erasing: the image of any non-empty word is never empty. Hence the action of a morphism $f$ on $\cA^*$ can be naturally extended to infinite words; that is, if $\bx = x_1x_2x_3 \cdots \in \cAw$, then $f(\bx) = f(x_1)f(x_2)f(x_3)\cdots$. 

In what follows, we will denote the composition of morphisms by juxtaposition as for concatenation of words.

\medskip
 
Episturmian morphisms are the compositions of the permutation morphisms (the morphisms $f$ such that $f(\cA)=\cA$) and the morphisms $L_a$ and $R_a$ where, for all $a \in \cA$:
\[
  L_a: \left\{\begin{array}{lll}
               a &\mapsto &a \\
               b &\mapsto &ab 
               \end{array}\right. , 
               \quad R_a: \left\{\begin{array}{lll}
               a &\mapsto &a \\
               b &\mapsto &ba    
               \end{array}\right. \quad \mbox{for all $b \ne a$ in $\cA$}.
\] 
Here we will work only on {\it pure} episturmian morphisms, i.e., morphisms obtained by composition of elements of  the sets:
\[
\mathcal{L}_\cA=\{L_a \mid a \in \cA\} \quad \mbox{and} \quad \mathcal{R}_\cA =\{R_a \mid a \in \cA\}.
\]

\begin{note}
In \cite{jJgP02epis}, the morphism $L_a$ (resp.~$R_a$) is denoted by $\psi_a$ (resp.~$\bar\psi_a$). We adopt the current notation to emphasize the action of $L_a$ (resp.~$R_a$) when applied to a word, which consists in placing an occurrence of the letter $a$ on the \underline{l}eft (resp.~\underline{r}ight) of each occurrence of any letter different from~$a$.
\end{note}

{\em Epistandard morphisms} are the morphisms obtained by concatenation of morphisms in $\mathcal{L}_\cA$ and permutations on $\cA$. Likewise, the  {\em pure episturmian morphisms} (resp.~{\em pure epistandard morphisms}) are the morphisms obtained by concatenation of morphisms in $\mathcal{L}_\cA \cup \mathcal{R}_\cA$ (resp.~in $\mathcal{L}_\cA$).
Note that the episturmian morphisms are exactly the {\em Sturmian morphisms} when $\cA$ is a $2$-letter alphabet.

All episturmian morphisms are injective on both finite and infinite words. The monoid of episturmian morphisms is \textit{left cancellative} (see \cite[Lem.~7.2]{gR03conj}) which means that for any episturmian morphisms $f, g, h$, if $fg = fh$ then $g = h$. Note that this fact, which is a by-product of the injectivity, can also be seen as a consequence of the invertibility of these morphisms (see \cite{aG06onst, eG06auto, gR03conj, zWyZ99some}). 

\subsection{Morphic decomposition of episturmian words} \label{SS:morphicDecomp}

Justin and Pirillo \cite{jJgP02epis} proved the following insightful characterizations of epistandard and episturmian words (see Theorem \ref{T:episturmian} below), which show that any episturmian word can be {\em infinitely decomposed} over the set of pure episturmian morphisms.

The statement of Theorem~\ref{T:episturmian} needs some extra definitions and notation.

First we define the following new alphabet, $\bar{\cA} = \{\bar{x} \mid x \in \cA \}$. A letter $\bar x$ is considered to be $x$ with {\em spin} $R$,  whilst $x$ itself has spin $L$. 
A finite or infinite word over 
$\cA \cup \bar{\cA}$ is called a \textit{spinned} word.
To ease the reading, we sometimes call a letter with spin $L$ (resp.~spin $R$) an $L$-spinned (resp.~$R$-spinned) letter. By extension, an $L$-spinned (resp.~$R$-spinned) word is a word having only letters with spin $L$ (resp.~spin $R$).

The {\em opposite} $\bar{w}$ of a finite or infinite spinned word $w$ is obtained from $w$ by exchanging all spins in $w$. For instance, if $v=ab \bar a$, then $\bar v=\bar a \bar b a$. When $v \in \cA^+$, then its opposite $\bar v \in \bar\cA^+$ is an $R$-spinned word and we set $\bar\empt = \empt$. Note that, given a finite or infinite word $w = w_1w_2\ldots$ over $\cA$, we sometimes denote $\breve w = {\breve w}_1{\breve w}_2\cdots$ any spinned word such that $\breve w_i = w_i$ if $\breve w_i$ has spin $L$ and $\breve w_i = \bar w_i$ if $\breve w_i$ has spin $R$. Such a word $\breve w$ is called a {\em spinned version} of $w$.

\begin{note} In Justin and Pirillo's original papers, spins are 0 and 1 instead of $L$ and $R$. It is convenient here to change this vision of the spins because of the relationship with episturmian morphisms, which we now recall.
\end{note}

For  $a \in \cA$, let $\mu_a = L_a$ and $\mu_{\bar a} = R_a$. This operator $\mu$ can be naturally extended (as done in \cite{jJgP02epis}) to a morphism from the free monoid $(\cA \cup \bar\cA)^*$ to a pure episturmian morphism: for a spinned finite word $\breve w = \breve w_1\ldots \breve w_n$ over $\cA \cup \bar\cA$, $\mu_{\breve w} = \mu_{\breve w_1}\ldots \mu_{\breve w_n}$ ($\mu_\varepsilon$ is the identity morphism). We say that the word $w$ {\em directs} or is a {\em directive word} of the morphism $\mu_w$. The following result extends the notion of directive words to infinite episturmian words.

\begin{thrm} {\rm \cite{jJgP02epis}} 
\label{T:episturmian}

\begin{enumerate} 
\item[$i)$] An infinite word $\bs \in \cAw$ is epistandard  if and only if there exist an infinite word  $\Delta = x_1x_2x_3\cdots$ over $\cA$ and an infinite sequence $(\bs^{(n)})_{n \geq 0}$ of infinite words such that $\bs^{(0)} = \bs$ and for all $n \geq 1$, $\bs^{(n-1)} = L_{x_n}(\bs^{(n)})$.

\item[$ii)$] An infinite word $\bt \in \cAw$ is episturmian if and only if there exist a spinned infinite word $\breve\Delta = \breve x_1 \breve x_2 \breve x_3 \cdots$ over $\cA \cup \bar\cA$ and an infinite sequence $(\bt^{(n)})_{n \geq 0}$ of recurrent infinite words such that    
$\bt^{(0)} = \bt$ and for all $n \geq 1$, $\bt^{(n-1)} = \mu_{\breve x_n}(\bt^{(n)})$. 
\end{enumerate}
\end{thrm}

For any epistandard word (resp.~episturmian word) $\bt$ and $L$-spinned infinite word $\Delta$ (resp.~spinned infinite word $\breve\Delta$) satisfying the conditions of the above theorem, we say that $\Delta$ (resp.~$\breve\Delta$) is a {\em (spinned) directive word} for $\bt$ or that $\bt$ is {\em directed by} $\Delta$ (resp.~$\breve\Delta$). 
Notice that this directive word is exactly the one that arises from the equivalent definition of epistandard words that uses {\em palindromic closure} \cite{xDjJgP01epis, aGjJ07epis, jJgP02epis} and, in the binary case, it is related to the continued fraction of the slope of the straight line represented by a standard word (see \cite{mL02alge}). 
It follows immediately from Theorem~\ref{T:episturmian} that, with the notation of case $ii)$, each $\bt^{(n)}$  is an episturmian word directed by $\breve x_{n+1} \breve x_{n+2} \cdots$

\pagebreak

The natural question: ``Does any spinned infinite word direct a unique episturmian word?'' is answered in \cite{jJgP02epis}:

\begin{prpstn}{\rm \cite[Prop.~3.11]{jJgP02epis}} 
\label{P:uniqueDirective}
\begin{enumerate}
\item  Any spinned infinite word $\breve \Delta$ having infinitely many $L$-spinned letters directs a unique episturmian word beginning with the left-most letter having spin $L$ in $\breve\Delta$. 
\item Any $R$-spinned infinite word $\breve \Delta$ directs exactly $|\eUlt(\Delta)|$ episturmian words.
\item Let $\breve \Delta$ be an $R$-spinned infinite word, and let $a$ be a letter such that $\bar a \in \eUlt(\breve\Delta)$. Then $\breve \Delta$ directs exactly one episturmian word starting with $a$. 
\end{enumerate} 
\end{prpstn}
\begin{note} 
In \cite{jJgP02epis}, item~3 was stated in the more general case where $\breve\Delta$ is ultimately $R$-spinned. In this case, $\breve \Delta$ still directs exactly one episturmian word for each letter $\bar a$ in $\Ult(\breve \Delta)$, but contrary to what is written in \cite{jJgP02epis}, nothing can be said on its first letter. 
\end{note}

As a consequence of the previous proposition and part $i)$ of Theorem~\ref{T:episturmian}, any $L$-spinned infinite word directs a unique epistandard word. 
The following important remark links the two parts of Theorem~\ref{T:episturmian}.

\begin{rmrk}\cite{jJgP02epis}
If $\breve\Delta$ is a spinned version of an $L$-spinned word $\Delta$ and if $\bt$ is  an episturmian word directed by $\breve\Delta$, then the set of factors of $\bt$ is exactly the set of factors of the epistandard word $\bs$ directed by $\Delta$.
\end{rmrk}

Moreover (with the same notation as in the previous remark):

\begin{rmrk} \label{R:periodic}
The episturmian word $\bt$ is periodic if and only if 
the epistandard word $\bs$ is periodic, and this holds if and only if
there is only one letter occurring infinitely often in $\Delta$, that is, $|\Ult(\Delta)| = 1$  (see \cite[Prop.~2.9]{jJgP02epis}). More precisely, a periodic episturmian word takes the form $(\mu_{\breve w}(x))^\omega$ for some finite spinned word $\breve w$ and letter $x$.
\end{rmrk}

\begin{note} Sturmian words are precisely the aperiodic episturmian words on a 2-letter alphabet.
\end{note}

When an episturmian word is aperiodic, we have the following fundamental link between the words $(\bt^{(n)})_{n \geq 0}$ and the spinned infinite word $\breve\Delta$ occurring in Theorem~\ref{T:episturmian}: if $a_n$ is the first letter of $\bt^{(n)}$, then $\mu_{\breve x_1 \ldots \breve x_n}(a_n)$ is a prefix of $\bt$ and the sequence $(\mu_{\breve x_1 \ldots \breve x_n}(a_n))_{n \geq 1}$ is not ultimately constant (since $\breve\Delta$ is not ultimately constant), then $\bt = \lim_{n \rightarrow \infty} \mu_{\breve x_1 \cdots \breve x_n}(a_n)$. This fact is a slight generalization of a result of Risley and Zamboni \cite[Prop.~III.7]{rRlZ00agen} on {\em S-adic representations} for characteristic Arnoux-Rauzy sequences.  See also the recent paper \cite{vBcHlZ06init} for S-adic representations of Sturmian words. Note that {\em $S$-adic dynamical systems} were introduced by Ferenczi \cite{sF99comp} as {\em minimal dynamical systems}  (e.g., see \cite{nP02subs})  generated by a finite number of substitutions. In the case of episturmian words, the notion itself is actually a reformulation of the well-known {\em Rauzy rules}, as studied in \cite{gR85mots}.

To anticipate next sections, let us also observe:

\begin{rmrk} \cite{jJgP02epis}
\label{R:spinnedVersions}
If an aperiodic episturmian word is directed by two spinned words $\Delta_1$ and $\Delta_2$, then $\Delta_1$ and $\Delta_2$ are spinned versions of a common $L$-spinned word.
\end{rmrk}

This is no longer true for periodic episturmian words; for instance $ab^\omega$ and $\bar{b}a^\omega$ direct the same episturmian word $(ab)^\omega = ababab\cdots$.

\section{Known results on directive-equivalent words} \label{S:directiveWords}

We have just seen an example of a periodic episturmian word that is directed by two different spinned infinite words. This situation holds also in the aperiodic case (see \cite{jJgP02epis, jJgP04epis}). For example, the {\em Tribonacci word} (or {\em Rauzy word}~\cite{gR82nomb}) is  directed by $(abc)^\omega$ and also by $(abc)^n\bar{a}\bar{b}\bar{c}(a\bar{b}\bar{c})^\omega$ for each $n\geq 0$, as well as infinitely many other spinned words. More generally, by \cite{jJgP02epis}, any epistandard word has a unique $L$-spinned directive word but also  has other directive words (see also  \cite{jJgP04epis} and Theorem~\ref{T:directSame}). 

We now consider in detail the following two questions: When do two finite spinned words direct a common episturmian morphism? When do two spinned infinite words direct a common unique episturmian word? We say that that two finite (resp.~infinite) spinned words are \emph{directive-equivalent} words if they direct a common episturmian morphism (resp.~a common episturmian word).

In Section~\ref{Ss:presentation}, we recall the characterizations of directive-equivalent finite spinned words. In Section~\ref{Ss:equivalencePast}, we recall known results about directive-equivalent infinite words. Section~\ref{S:equivalence} will present a new characterization of these words.

\subsection{\label{Ss:presentation}Finite directive-equivalent words: presentation versus block-equivalence}

Generalizing a study of the monoid of Sturmian morphisms by Séébold \cite{See1991}, the third author \cite{gR03conj} answered the question: ``When do two spinned finite words direct a common episturmian morphism?'' by giving a presentation of the monoid of episturmian morphisms. This result was reformulated in \cite{gR03lynd} using another set of generators and it was independently and differently treated in \cite{jJgP04epis}. As a direct consequence, one can see that the monoid of pure epistandard morphisms is a free monoid and one can obtain the following presentation of the monoid of pure episturmian morphisms:

\begin{thrm}\label{T:presentation}{\rm (direct consequence of \cite[Prop.~6.5]{gR03lynd}; reformulation of \cite[Th.~2.2]{jJgP04epis})}

The monoid of pure episturmian morphisms with $\{L_{\alpha}, R_{\alpha} \mid \alpha\in \cA\}$ as a set of generators has the following presentation:
$$R_{a_1} R_{a_2} \ldots R_{a_k} L_{a_1} = L_{a_1} L_{a_2} \ldots L_{a_k} R_{a_1}$$
where $k \geq 1$ is an integer and $a_1, \ldots ,a_k \in \cA$ with $a_1 \neq a_i$ for all $i$, $2 \leq i \leq k$. 
\end{thrm}

This result means that two different compositions of morphisms in ${\cal L}_{\cal A} \cup  {\cal R}_{\cal A}$ yield a common pure episturmian morphism if and only if one composition can be deduced from the other one in a rewriting system, called the {\em block-equivalence} in \cite{jJgP04epis}. 
Although Theorem~\ref{T:presentation} allows us to show that many properties of episturmian words are linked to properties of episturmian morphisms, it will be convenient for us to have in mind the block-equivalence that we now recall.

A word of the form $xvx$, where $x \in \cA$ and $v \in (\cA\setminus\{x\})^*$, is called a ($x$-based) {\em block}. A ($x$-based) {\em block-transformation} is the replacement in a spinned word of an occurrence of $xv\bar x$ (where $xvx$ is a block) by $\bar x \bar v x$ or vice-versa. Two finite spinned words $w$, $w'$ are said to be {\em block-equivalent} if  we can pass from one to the other by a (possibly empty) chain of block-transformations, in which case we write $w \equiv w'$. For example, $\bar b\bar a b \bar c b \bar a \bar c$ and  $b a bc\bar b\bar a \bar c$ are block-equivalent because $\bar b\bar a b \bar c b \bar a \bar c \rightarrow ba\bar b \bar c b\bar a \bar c  \rightarrow b a bc\bar b\bar a \bar c$ and vice-versa. 
The block-equivalence is an equivalence relation over spinned words, and moreover one can observe that if $w \equiv w'$ then $w$ and $w'$ are spinned versions of a common word over $\cA$.

\medskip

Theorem~\ref{T:presentation} can be reformulated in terms of block-equivalence:

\medskip
\noindent
\textbf{Theorem~\ref{T:presentation}.} 
\textit{Let $w$, $w'$ be two spinned words over $\cA \cup \bar\cA$. Then $\mu_{w} = \mu_{w'}$ if and only if $w \equiv w'$. 
}

\subsection{\label{Ss:equivalencePast}Infinite directive-equivalent words: previous results}

 The question: ``When do two spinned infinite words direct a common unique episturmian word?'' was tackled  by Justin and Pirillo in \cite{jJgP04epis} for {\em bi-infinite episturmian words}, that is, episturmian words with letters indexed by $\ZZ$ (and not by $\NN$ as considered until now). Let us recall relations between right-infinite episturmian words and bi-infinite episturmian words (see \cite[p.~332]{jJgP04epis} and \cite{aGjJ07epis} for more details). 

 First we observe that a right-infinite episturmian word $\bt$ can be prolonged infinitely to the left with the same set of factors. Note also that the definition of episturmian words considered in Section~\ref{Ss:defEpisturmian} (using reversal and special factors) can be extended to bi-infinite words (see \cite{jJgP04epis}). 
Furthermore, the characterization (Theorem~\ref{T:episturmian}) of right-infinite episturmian words by a sequence $(\bt^{(i)})_{i\geq0}$ extends to bi-infinite episturmian words, with all the $\bt^{(i)}$ now bi-infinite episturmian words. That is, as for right-infinite episturmian words, we have bi-infinite words of the form $\bl^{(i)}.\br^{(i)}$ where $\bl^{(i)}$ is a left-infinite episturmian word and $\br^{(i)}$ is a right-infinite episturmian word. Moreover, if the bi-infinite episturmian word $\bb = \bl.\br$ is directed by $\breve\Delta$ with associated bi-infinite episturmian words $\bb^{(i)} = \bl^{(i)}.\br^{(i)}$, then $\br$ is directed by $\breve\Delta$ with associated right-infinite episturmian words $\br^{(i)}$.

As a consequence of what precedes, Justin and Pirillo's results about spinned words directing a common bi-infinite episturmian word are still valid for words directing a common (right-infinite) episturmian word. We summarize now these results, which will be helpful for the proof of our main theorem (Theorem~\ref{T:directSame}, to follow).

\medskip
First of all, Justin and Pirillo characterized pairs of words directing a common episturmian word in the case
of {\em wavy} directive words, that is, spinned infinite words containing infinitely many $L$-spinned letters and infinitely many $R$-spinned letters. This characterization uses the following extension of the block-equivalence $\equiv$ for infinite words.

Let $\Delta_1$, $\Delta_2$ be spinned versions of $\Delta$. We write $\Delta_1 \rightsquigarrow \Delta_2$ if there exist infinitely many prefixes ${f}_i$ of $\Delta_1$ and ${g}_i$ of $\Delta_2$  with the $g_i$ of strictly increasing lengths, and such that, for all $i$, $|g_i| \leq |f_i|$ and  ${f}_i \equiv {g}_i {c}_i$ for a suitable spinned word ${c}_i$. 
Infinite words $\Delta_1$ and $\Delta_2$
are said to be \textit{block-equivalent} (denoted by $\Delta_1 \equiv \Delta_2$) if $\Delta_1 \rightsquigarrow \Delta_2$ and $\Delta_2 \rightsquigarrow \Delta_1$.

\begin{thrm}{\rm \cite[Th.~3.4, Cor.~3.5]{jJgP04epis}}
\label{T:JP_theo3.4}
Let $\Delta_1$ and $\Delta_2$ be wavy spinned versions of $\Delta \in A^\omega$ with $|\eUlt(\Delta)|> 1$.
Then $\Delta_1$ and $\Delta_2$ direct a common (unique) episturmian word if and only if $\Delta_1 \equiv \Delta_2$.

Moreover when $\Delta_1$ and $\Delta_2$ do not have any common prefix modulo $\equiv$, and when there exists a letter $x$ such that $\Delta_1$ and $\Delta_2$ begin with $x$ and $\bar{x}$ respectively, if $\Delta_1 \equiv \Delta_2$, then 
$\Delta_1 = x \prod_{n \geq 1} v_n {\breve x}_n$, $\Delta_2 = {\bar x} \prod_{n \geq 1} {\bar v}_n {\hat x}_n$ for an $L$-spinned letter $x$, a sequence $(v_n)_{n \geq 1}$ of $x$-free $L$-spinned words, and  sequences of spinned letters $({\breve x}_n)_{n \geq 1}$, $({\hat x}_n)_{n \geq 1}$ in $\{x, \bar x\}$ such that $({\breve x}_n)_{n \geq 1}$ contains infinitely many times the $R$-spinned letter $\bar{x}$, and $({\hat x}_n)_{n \geq 1}$ contains infinitely many times the $L$-spinned letter ${x}$.
\end{thrm}

The relation $\rightsquigarrow$ (and hence the block-equivalence $\equiv$ for infinite words) is rather intricate to understand. So in some way the forms of $\Delta_1$ and $\Delta_2$ at the end of Theorem~\ref{T:JP_theo3.4} are, although technical, easier to understand. Theorem~\ref{T:directSame}, which refines the end of the previous result and proves the converse, describes all possible forms for pairs of directive-equivalent words without any use of notations $\rightsquigarrow$ and $\equiv$.

\medskip
When one of the two considered directive words is not wavy, Justin and Pirillo established:

\begin{prpstn}{\rm \cite[Prop.~3.6]{jJgP04epis}}
\label{P:JP_Prop3.6}
Let $\Delta_1$ and $\Delta_2$ be spinned versions of a common word such that $\Delta_1$ is wavy and letters of $\Delta_2$ are ultimately of spin $L$ (resp.~ultimately of spin $R$). If $\Delta_1$ and $\Delta_2$ are directive-equivalent, then $\Delta_1 \rightsquigarrow
\Delta_2$. Moreover there exist spinned words $w_1$, $w_2$, an $L$-spinned letter $x$, and $L$-spinned $x$-free words $(v_i)_{i \geq 1}$ 
such that $\mu_{w_1} = \mu_{w_2}$, $\Delta_1 = w_1 \bar{x}\prod_{i \geq 1} \bar{v}_i x$
 and $\Delta_2 = w_2x\prod_{i \geq 1} v_i x$
(resp.~$\Delta_1 = w_1 {x}\prod_{i \geq 1} {v}_i \bar{x}$
 and $\Delta_2 = w_2\bar{x}\prod_{i \geq 1} \bar{v}_i \bar{x}$).
\end{prpstn}

\medskip
With the next two results, they considered the remaining cases of words directing aperiodic episturmian words. In the first one, the spins of the letters in each of the two directive words are ultimately $L$  or ultimately $R$. The second result shows that if one of the directive words has the spins of its letters ultimately $L$ (resp.~ultimately $R$), then the other directive word cannot have the spins of its letters ultimately $R$ (resp.~ultimately $L$). 
\begin{prpstn}{\rm \cite[Prop.~3.7]{jJgP04epis}}
\label{P:JP_Prop3.7}
Let $\Delta_1$ and $\Delta_2$ be spinned versions of a common word $\Delta \in \cAw$ with $|\eUlt(\Delta)| > 1$. If there exist spinned words $w_1, w_2$ and an $L$-spinned infinite word $\Delta'$ such that $\Delta_1=w_1\Delta'$ and $\Delta_2=w_2\Delta'$ 
(resp.~$\Delta_1=w_1\bar{\Delta}'$ and $\Delta_2=w_2\bar{\Delta}'$), then $\Delta_1$, $\Delta_2$ are directive-equivalent if and only if $\mu_{w_1} = \mu_{w_2}$.
\end{prpstn}

\begin{prpstn}{\rm \cite[Prop.~3.9]{jJgP04epis}}
\label{P:JP_Prop3.9} 
Let $\Delta$ be an $L$-spinned infinite word. Then $\Delta$ and $\bar{\Delta}$ do not direct a common right-infinite episturmian word.
\end{prpstn}

Actually the previous statement is a corollary of Proposition~3.9 in \cite{jJgP04epis} which considers more generally words directing episturmian words differing only by a shift.

\medskip

Justin and Pirillo also discussed in \cite{jJgP04epis} the periodic case and proved:
\begin{prpstn}{\rm \cite[Prop.~3.10]{jJgP04epis}}
\label{P:JP_Prop3.10} Suppose that $\Delta_1 = \breve{w}\breve{y}a^\omega$ and
$\Delta_2 = \hat{w}\hat{y}\bar{a}^\omega$, where $\breve{w}$ and $\hat{w}$ (resp.~$\breve{y}$ and $\hat{y}$) are spinned versions of a common word and $a$ is an $L$-spinned letter.
Then $\Delta_1$ and $\Delta_2$ are directive-equivalent if and only if there exist sequences of letters $(\breve{a}_n)_{n\geq 1}$ and $(\hat{a}_n)_{n\geq 1}$ such that $\breve{w}\breve{y}\prod_{n\geq 1} \breve{a}_n \equiv
\hat{w}\hat{y}\prod_{n\geq 1} \hat{a}_n$.
\end{prpstn}

We will see in Theorem~\ref{T:directSame} that other cases can occur for periodic episturmian words.

\section{\label{S:equivalence}Directive-equivalent words: a characterization}

As shown in the previous section, Justin and Pirillo provided quite complete results about directive-equivalent infinite words. Nevertheless they did not systematically provide the relative forms of two directive-equivalent words. The following characterization does it, moreover without the use of relations $\rightsquigarrow$ and $\equiv$. This result also fully solves the periodic case, which was only partially solved in~\cite{jJgP04epis}.


 \begin{thrm}\label{T:directSame} 
Given two spinned infinite words $\Delta_1$ and $\Delta_2$, the following assertions are equivalent.
\begin{description}
\item{i)} $\Delta_1$ and $\Delta_2$ direct a common right-infinite episturmian word;
\item{ii)} $\Delta_1$ and $\Delta_2$ direct a common bi-infinite episturmian word;
\item{iii)} One of the following cases holds for some $i, j$ such that $\{i, j\} = \{1, 2\}$:
\begin{enumerate}
\item \label{Ti:1} $\Delta_i = \prod_{n \geq 1} v_n$, $\Delta_j = \prod_{n \geq 1} z_n$ where $(v_n)_{n \geq 1}, (z_n)_{n \geq 1}$ are spinned words such that $\mu_{v_n} = \mu_{z_n}$ for all $n \geq 1$;

\item \label{Ti:2} $\Delta_i = {w} x \prod_{n \geq 1} v_n {\breve x}_n$, $\Delta_j = {w'} {\bar x} \prod_{n \geq 1} {\bar v}_n {\hat x}_n$ where ${w}$, ${w'}$ are spinned words such that $\mu_{w} = \mu_{w'}$, $x$ is an $L$-spinned letter, $(v_n)_{n \geq 1}$ is a sequence of non-empty $x$-free $L$-spinned words, and $({\breve x}_n)_{n \geq 1}$, $({\hat x}_n)_{n \geq 1}$ are sequences of non-empty spinned words over $\{x, \bar x\}$ such that, for all $n \geq 1$, $|{\breve x}_n| = |{\hat x}_n|$ and  $|{\breve x}_n|_x = |{\hat x}_n|_x$;

\item \label{Ti:4} $\Delta_1 = w \bx$ and $\Delta_2 = w'\by$ where $w$, $w'$ are spinned words, $x$ and $y$ are letters,  and  $\bx \in \{x, \bar x\}^\omega$, $\by \in \{y, \bar y\}^\omega$ are spinned infinite words such that $\mu_{w}(x) = \mu_{w'}(y)$.

\end{enumerate} 
\end{description}
 \end{thrm}

\begin{note}
For $a, b, c$ three different letters in $\cA$, the spinned infinite words $\Delta_1 = a(bc{\bar a})^\omega$ and $\Delta_2 = {\bar a}({\bar b}{\bar c}{\bar a})^\omega$ direct a common episturmian word that starts  with the letter $a$.
Indeed, these two directive words fulfill item 2 of Theorem~\ref{T:directSame} with $w = w' = \varepsilon$, $x = a$, and for all $n$, $v_n = bc$ and $\breve{x}_n = \hat{x}_n = \bar{a}$. Moreover the fact that $\Delta_1$ starts with the $L$-spinned letter $a$ shows that the word it directs starts with $a$. 
Similarly $\Delta_1' = {\bar a}b(ca{\bar b})^\omega$ and $\Delta_2' = {\bar a}{\bar b}({\bar c}{\bar a}{\bar b})^\omega$ 
 direct a common episturmian word starting with the letter $b$.
Since $\Delta_2 = \Delta_2'$,  
this shows that the relation ``direct a common episturmian word'' over spinned infinite word is not an equivalence relation. 
\end{note}

Items \ref{Ti:2} and \ref{Ti:4} of Theorem~\ref{T:directSame} show that any episturmian word is directed by a spinned infinite word having infinitely many $L$-spinned letters, but also by a spinned word having both infinitely many $L$-spinned letters and infinitely many $R$-spinned letters (i.e., a wavy word). To emphasize the importance of these facts, let us recall from Proposition~\ref{P:uniqueDirective} that if $\breve\Delta$ is a spinned infinite word over $\cA \cup \bar\cA$ with infinitely many $L$-spinned letters, then there exists a unique episturmian word $\bt$ directed by $\breve\Delta$. Unicity comes from the fact that the first letter of $\bt$ is fixed by the first $L$-spinned letter in $\breve\Delta$.

Before proving Theorem~\ref{T:directSame}, let us make two more remarks.

\begin{rmrk}
\label{R:AboutDirectSame3}
In items \ref{Ti:1} and \ref{Ti:2} of Theorem~\ref{T:directSame}, the two considered directive words are spinned versions of a common $L$-spinned word. This does not hold in item \ref{Ti:4}, which deals only with periodic episturmian words. This is consistent with Remark~\ref{R:spinnedVersions}.  As an example of item~\ref{Ti:4}, one can consider the word $(ab)^\omega = L_a(b^\omega) = R_b(a^\omega)$ which, as already said at the end of Section~\ref{SS:morphicDecomp}, is directed by $a{b}^\omega$ and by $\bar{b}a^\omega$ ($L_a(b) = ab = R_b(a)$). Note also that $(ab)^\omega$ is directed by $(a\bar b)^\omega$, underlining the fact that $x$ and $y$ can be equal in item~\ref{Ti:4} of Theorem~\ref{T:directSame}.
\end{rmrk}

\begin{rmrk} \label{R:AboutDirectSame4} If an episturmian word $\bt$ has two directive words satisfying items \ref{Ti:2} or \ref{Ti:4}, then $\bt$ has infinitely many directive words. Indeed, if item~\ref{Ti:2} is satisfied and $\bar x$ occurs in ${\breve x}_p$ ($p \geq 1$), then by Theorem \ref{T:presentation},
$x \left(\prod_{k=1}^{p-1} {v}_n{\breve x}_n\right){v}_p{\breve x}_p'\bar{x} \equiv
\bar x \left(\prod_{k=1}^{p-1} {\bar v}_n{\breve x}_n\right){\bar v}_p{\breve x}_p'x$  where ${\breve x}_p'$ is such that ${\breve x}_p \equiv \bar x{\breve x}_p'$. Thus
 $\bt$ is also directed by  $w\bar x \left(\prod_{k=1}^{p-1} {\bar v}_n{\breve x}_n\right){\bar v}_p{\breve x}_p'x\prod_{n\geq p+1} {v}_n{\breve x}_n$. 
Similarly if item~\ref{Ti:2} is satisfied and $x$ occurs in ${\breve x}_p$ ($p \geq 1$), then  $\bt$ is also directed by  $w'x \left(\prod_{k=1}^{p-1} {v}_n{\hat x}_n\right)v_p{\hat x}_p'\bar{x}\prod_{n\geq p+1} {\bar v}_n{\hat x}_n$ where ${\hat x}_p'$ is such that ${\hat x}_p \equiv x{\hat x}_p'$.
If item~\ref{Ti:4} is satisfied, then $\bt$ is periodic and directed by $w\bx$ where $\bx$ is any spinned version of $x^\omega$.
\end{rmrk}

The rest of this section is dedicated to the proof of Theorem~\ref{T:directSame}.

\begin{proof}[Proof of Theorem $\ref{T:directSame}$] 
We have $i) \Leftrightarrow ii)$ by the remarks on bi-infinite words at the beginning of Section~\ref{Ss:equivalencePast}.

\medskip

\noindent $iii) \Rightarrow i)$.  Assume first that $\Delta_1 = \prod_{n \geq 1} v_n$ and $\Delta_2 = \prod_{n \geq 1} z_n$ for spinned words $(v_n)_{n \geq 1}, (z_n)_{n \geq 1}$ such that $\mu_{v_n} = \mu_{z_n}$ for all $n \geq 1$. From the latter equality and Theorem~\ref{T:presentation}, $\Delta_1$ has infinitely many $L$-spinned letters if and only if $\Delta_2$ has infinitely many $L$-spinned letters. 

Let us first consider the case when both $\Delta_1$ and $\Delta_2$ have infinitely many $L$-spinned letters. Without loss of generality we can assume that for all $n \geq 1$, $v_n$ and $z_n$ contain at least one $L$-spinned letter. 
Now we need to define some more notations. Let $\bt_1$ and $\bt_2$ be the episturmian words directed by $\Delta_1$ and $\Delta_2$, respectively (these episturmian words exist and are unique by Proposition~\ref{P:uniqueDirective}). For $n \geq 0$, let $\bt_1^{(n)}$ and $\bt_2^{(n)}$ be the episturmian words as in $ii)$ of Theorem~\ref{T:episturmian} and let $a_n$ and $b_n$ be their respective  first letters. 
Finally, for $n \geq 1$, set $p_n = \prod_{i = 1}^n v_i$ and $q_n = \prod_{i = 1}^n z_i$. The words $\mu_{p_n}(a_{|p_n|})$ (resp.~$\mu_{q_n}(b_{|q_n|})$) are prefixes of $\bt_1$ (resp.~of $\bt_2$). 
The letter $a_{|p_n|}$ (resp.~$b_{|q_n|}$) is the first letter of $\mu_{v_{n+1}}(\bt_1^{(m)})$ (resp.~$\mu_{z_{n+1}}(\bt_2^{(m)})$) with $m = \sum_{i = 1}^{n+1} |v_i| = \sum_{i = 1}^{n+1} |z_i|$. Since $v_{n+1}$ (resp.~$z_{n+1}$) contains at least one $L$-spinned letter, $a_{|p_n|}$ (resp.~$b_{|q_n|}$) is the first letter of $\mu_{v_{n+1}}(w)$ (resp.~$\mu_{z_{n+1}}(w)$) for any word $w$. From $\mu_{v_{n+1}} = \mu_{z_{n+1}}$, we have $a_{|p_n|} = b_{|q_n|}$ and so $\mu_{p_n}(a_{|p_n|}) = \mu_{q_n}(b_{|q_n|})$ for all $n \geq 1$. If the sequence $(\mu_{p_n}(a_{|p_n|}))_{n \geq 1}$ is not ultimately constant, then from $\bt_1 = \lim_{n \to \infty}\mu_{p_n}(a_{|p_n|})$ and  $\bt_2 = \lim_{n \to \infty} \mu_{q_n}(b_{|q_n|})$, we deduce that $\bt_1 = \bt_2$. If $(\mu_{p_n}(a_{|p_n|}))_{n \geq 1}$ is ultimately constant, then necessarily there exists a letter $a$ and an integer $m$ such that for all $n > m$, $v_n$ and $z_n$ belong to $\{a\}^*$. Then $\bt_1 = \mu_{v_1\ldots v_m}(a^\omega) = \mu_{z_1\ldots z_m}(a^\omega) = \bt_2$. 

Now, with the same notations as in the above case, we consider the case when 
the letters of $\Delta_1$ and $\Delta_2$ are ultimately $R$-spinned. By Theorem~\ref{T:presentation}, any equality $\mu_v = \mu_z$ (for some different spinned words $v$ and $z$) implies that $v$ and $z$ both contain at least one $L$-spinned letter  and one $R$-spinned letter. Hence, in our current case, 
there exists an integer $m$ such that  $v_n = z_n$ for all $n > m$.   
Let $\bt$ be an episturmian word directed by $\prod_{n > m} v_n  = \prod_{n > m} z_n$ (such an episturmian word exists by Proposition~\ref{P:uniqueDirective}). Then $\mu_{p_m}(\bt) = \mu_{q_m}(\bt)$ and $\Delta_1$ and  $\Delta_2$ are directive-equivalent.

\medskip

Now consider item~2 of part $iii)$. 
We assume that $\Delta_1 = {w} x \prod_{n \geq 1} v_n {\breve x}_n$ and $\Delta_2 = {w'} {\bar x} \prod_{n \geq 1} {\bar v}_n {\hat x}_n$ where ${w}$, ${w'}$ are spinned words such that $\mu_{w} = \mu_{w'}$, $x$ is an $L$-spinned letter, $(v_n)_{n \geq 1}$ is a sequence of non-empty $x$-free $L$-spinned words, and $({\breve x}_n)_{n \geq 1}$, $({\hat x}_n)_{n \geq 1}$ are non-empty spinned words over $\{x, \bar x\}$ such that, for all $n \geq 1$, $|{\breve x}_n| = |{\hat x}_n|$ and  $|{\breve x}_n|_x = |{\hat x}_n|_x$. By injectivity of the morphisms $\mu_{w} = \mu_{w'}$, $\Delta_1$ and $\Delta_2$ are directive-equivalent if and only if $w^{-1}\Delta_1$ and $w'^{-1}\Delta_2$ are directive-equivalent. So, from now on, we assume without loss of generality that $w = w' = \varepsilon$.

By Proposition~\ref{P:uniqueDirective}, there exist unique episturmian words $\bt_1$ and $\bt_2$ starting with $x$ directed by the respective words $\Delta_1$ and $\Delta_2$ (observe that if ${\hat x}_n \in {\bar x}^+$ for all $n\geq 1$, then $\bar x \in \Ult(\Delta_2)$). For $i \geq 1$, let $\Delta_1^{(i)} = x\prod_{n \geq i} v_n {\breve x}_n$
and $\Delta_2^{(i)} = \bar x\prod_{n \geq i} {\bar v}_n {\hat x}_n$ and let $\bt_1^{[i]}$ and $\bt_2^{[i]}$ be the words beginning with $x$ and directed by the respective words $\Delta_1^{(i)}$ and $\Delta_2^{(i)}$. (The episturmian words $\bt_1^{[i]}$ and $\bt_2^{[i]}$ exist by Proposition~ \ref{P:uniqueDirective}.) For $i \geq 1$ we also define $\alpha_i := |{\breve x}_i|_x = |{\hat x}_i|_x$ and $\beta_i := |{\breve x}_i|_{\bar x} = |{\hat x}_i|_{\bar x}$.

Assume first that $\alpha_i \neq 0$. Then $\bar x {\bar v}_i {\hat x}_i \equiv \bar x {\bar v}_i x x^{\alpha_i-1} {\bar x}^{\beta_i} \equiv x v_i \bar x x^{\alpha_i-1} {\bar x}^{\beta_i} \equiv 
x v_i x^{\alpha_i-1} {\bar x}^{\beta_i}\bar x$ and $x v_i {\breve x}_i \equiv x v_i x^{\alpha_i-1} {\bar x}^{\beta_i}x$. Let $p_i = x v_i x^{\alpha_i-1}{\bar x}^{\beta_i}$. From what precedes we deduce that $\Delta_1^{(i)}$ and $p_i\Delta_1^{(i+1)}$ are directive-equivalent, as $\Delta_2^{(i)}$ and $p_i\Delta_2^{(i+1)}$ are directive-equivalent. By the choice of words $\bt_1^{[i]}$ and $\bt_2^{[i]}$, we deduce that $\bt_1^{[i]} = \mu_{p_i}(\bt_1^{[i+1]})$ and $\bt_2^{[i]} = \mu_{p_i}(\bt_2^{[i+1]})$ and each of these words starts with $\mu_{p_i}(x)$.

Now let us consider the case when $\alpha_i = 0$. Then ${\breve x}_i = {\hat x}_i = {\bar x}^{\beta_i}$.
We have $x v_i {\breve x}_i \equiv {\bar x}{\bar v}_i {\bar x}^{\beta_i-1}x$ and
${\bar x}{\bar v}_i{\hat x}_i = {\bar x}{\bar v}_i {\bar x}^{\beta_i-1}\bar x$.
Taking $p_i = {\bar x}{\bar v}_i {\bar x}^{\beta_i-1}$, we reach the same conclusion as in the case when $\alpha_i \neq 0$.

It follows from what precedes that $\bt_1$ and $\bt_2$ both start with $\mu_{p_1\ldots p_i}(x)$ for all $i \geq 1$. Since $v_i \neq \varepsilon$, the sequence $(\mu_{p_1\ldots p_i}(x))_{i \geq 1}$ is not ultimately constant; whence $\bt_1 = \bt_2 = \lim_{i \to \infty} \mu_{p_1\ldots p_i}(x)$.

\medskip

Lastly, assume that $\Delta_1 = w \bx$ and $\Delta_2 = w' \by$ for some spinned words $w$, $w'$, some letters $x$ and $y$,  and some spinned infinite words $\bx \in \{x, \bar x\}^\omega$, $\by \in \{y, \bar y\}^\omega$ such that $\mu_{w}(x) = \mu_{w'}(y)$. The word $\Delta_1$ (resp.~$\Delta_2$) directs the episturmian word $\mu_{w}(x^\omega) = (\mu_w(x))^\omega$ (resp.~$\mu_{w'}(y^\omega) = (\mu_{w'}(y))^\omega$). Hence $\Delta_1$ and $\Delta_2$ are directive-equivalent. 

\bigskip

\noindent $i) \Rightarrow iii)$. Suppose $\Delta_1$ and $\Delta_2$ direct a common (right-infinite) episturmian word $\bt$. Let us first assume that $\bt$ is aperiodic. Then, by Remark~\ref{R:spinnedVersions}, $\Delta_1$ and $\Delta_2$ are spinned versions of a common infinite word $\Delta \in \cAw$. 
We now show that item 1 or item 2 holds using results of Justin and Pirillo in \cite{jJgP04epis}.  
\medskip

First consider the case when both $\Delta_1$ and $\Delta_2$ are wavy.  
Suppose there exist a sequence of prefixes $(p_n)_{n\geq 0}$ of $\Delta_1$ and a sequence of prefixes 
$(p_n')_{n \geq 0}$ of $\Delta_2$ such that for all $n \geq 0$, 
$\mu_{p_n} = \mu_{p_n'}$. Without loss of generality we can assume that $p_0 = p_0' = \varepsilon$  and the sequence $(|p_n|)_{n\geq 0}$ is strictly increasing. For $n \geq 1$, let $v_n$, $z_n$ be such that $p_n = p_{n-1}v_n$, $p_n' = p_{n-1}'z_n$; that is $\Delta_1 = \prod_{n \geq 1} v_n$ and
$\Delta_2 = \prod_{n \geq 1} z_n$. Let us prove by induction that  $\mu_{v_n} = \mu_{z_n}$ for all $n \geq 1$. First $\mu_{v_1} = \mu_{p_1} = \mu_{p_1'} = \mu_{z_1}$. For $n \geq 2$, since $\mu_{p_n} = \mu_{p_{n-1}}\mu_{v_n}$, $\mu_{p_n'} = \mu_{p_{n-1}'}\mu_{z_n}$, $\mu_{p_n} = \mu_{p_n'}$ and $\mu_{p_{n-1}} = \mu_{p_{n-1}'}$, we have $\mu_{p_{n-1}}\mu_{v_n} = \mu_{p_{n-1}}\mu_{z_n}$ and so 
$\mu_{v_n} = \mu_{z_n}$ by left cancellativity of the monoid of episturmian morphisms. 
So item~\ref{Ti:1} is satisfied in this case.

Now assume that previous sequences $(p_n)_{n\geq 0}$ and 
$(p_n')_{n \geq 0}$ do not exist. Let $w$ and $w'$ be the longest prefixes of the respective spinned words $\Delta_1$ and $\Delta_2$ such that $\mu_{w} = \mu_{w'}$. Further, let  
$\Delta_1'$ and $\Delta_2'$ be the spinned words such that $\Delta_1  = {w}\Delta_1'$ and
$\Delta_2  = {w'}\Delta_2'$. Then, by injectivity of $\mu_w$, the words $\Delta_1'$ and $\Delta_2'$ are directive-equivalent and have no prefixes with equal images by $\mu$. 

By Theorem~\ref{T:JP_theo3.4}, there exists a letter $x$ in $\cA$, a sequence of non-empty $x$-free words $(v_n)_{n \geq 1}$ over $\cA$, and two sequences of non-empty words $(\breve x_n)_{n \geq 1}$, $(\hat x_n)_{n \geq 1}$ over $\{x, {\bar x}\}$ such that $\Delta_i' = x\prod_{n \geq 1}v_n {\breve x}_n$ and $\Delta_j' = {\bar x}\prod_{n \geq 1}{\bar v}_n {\hat x}_n$ for some integers $i, j$ such that $\{i, j\} = \{1, 2\}$. 
We have to prove that for all $n \geq 1$, $|{\breve x}_n| = |{\hat x}_n|$ and  $|{\breve x}_n|_x = |{\hat x}_n|_x$. We use induction on $n$ and prove also that for all $n \geq 0$,
the words $\Delta_i^{(n+1)} = x\prod_{m \geq n+1}v_m {\breve x}_m$ and $\Delta_j^{(n+1)} = {\bar x}\prod_{m \geq n+1}{\bar v}_m {\hat x}_m$ are directive-equivalent. 

Let $n \geq 1$ be an integer. By definition of $\Delta_i^{(1)} = \Delta_i'$ and  $\Delta_j^{(1)} = \Delta_j'$ (when $n = 1$) and by the induction hypothesis (when $n \geq 2$), we know that the words $\Delta_i^{(n)} = x\prod_{m \geq n}v_m {\breve x}_m$ and $\Delta_j^{(n)} = {\bar x}\prod_{m \geq n}{\bar v}_m {\hat x}_m$ are directive-equivalent.

Assume first that ${\hat x}_n$ contains at least one occurrence of $x$. Then, with $\alpha_n = |{\hat x_n}|_x$ and $\beta_n = |{\hat x_n}|_{\bar x}$, we have
$\bar x{\bar v}_n {\hat x}_n \equiv \bar x{\bar v}_n x x^{\alpha_n-1}{\bar x}^{\beta_n} \equiv x v_n x^{\alpha_n-1}{\bar x}^{\beta_n} \bar x$. By injectivity of the morphism $\mu_{xv_n}$ we deduce that the words ${\breve x_n}\prod_{m \geq n+1} v_m {\breve x}_m = {\breve x_n}v_{n+1}{\breve x_{n+1}}\prod_{m \geq n+2} v_m {\breve x}_m$ and
$x^{\alpha_n-1}{\bar x}^{\beta_n} \bar{x} \prod_{m \geq n+1} {\bar v}_m {\hat x}_m = x^{\alpha_n-1}{\bar x}^{\beta_n+1} \bar{v}_{n+1}{\hat x_{n+1}}\prod_{m \geq n+2} {\bar v}_m {\hat x}_m$ direct a common episturmian word~$\bt_n$. The word $v_{n+1}$ is not empty. Let $c$ be its first letter, 
let $D = c^{-1}v_{n+1}\breve{x}_{n+1}\prod_{m \geq n+2}v_m\breve{x}_m$ and let
$D' = (\bar{c})^{-1}\bar{v}_{n+1}\hat{x}_{n+1}\prod_{m \geq n+2}\bar{v}_m\hat{x}_m$: 
the word $\bt_n$ is directed by $\breve{x}_ncD$ and by $x^{\alpha_n-1}\bar{x}^{\beta_n+1}\bar{c}D'$. 
Since $\Delta_j$ is wavy, $D'$ is also wavy. So $x$ occurs in $D'$ (among the $\hat{x}_n$) and the word directed by $D'$ starts with $x$. 
Consequently $\bt_n$ starts with $\mu_{x^{\alpha_n-1}\bar{x}^{\beta_n+1}\bar{c}}(x) = x^{\alpha_n}cx^{\beta_{n+1}}$. 
The words $v_m$ are non-empty, thus there exists a letter $d\neq x$ that occurs in the word directed by $D'$. Consequently $cx^{\alpha_n+\beta_n}d$ is the smallest factor of $\bt_n$ belonging to $c\{x\}^*d$. Since $\bt_n$ is also directed by $\breve{x}_ncD$, it follows that $\bt_n$ starts with $x^{|\breve{x}_n|_x}c$ and the smallest factor of $\bt_n$ belonging to $c\{x\}^*d$ is $cx^{|\breve{x}_n|}d$. Hence 
$|{\breve x}_n|_x = \alpha_n = |{\hat x}_n|_x$ and 
$|{\breve x}_n| = \alpha_n + \beta_n = |{\hat x}_n|$.
Consequently ${\breve x_n} \equiv x^{\alpha_n}{\bar x}^{\beta_n} \equiv x^{\alpha_n-1}{\bar x}^{\beta_n}x$. The injectivity of the morphism $\mu_{x^{\alpha_n-1}{\bar x}^{\beta_n}}$ implies that $\Delta_i^{(n+1)} = x\prod_{m \geq n+1}v_m {\breve x}_m$ and $\Delta_j^{(n+1)} = {\bar x}\prod_{m \geq n+1}{\bar v}_m {\hat x}_m$ are directive-equivalent. 

When ${\breve x}_n$ contains at least one occurrence of ${\bar x}$, we similarly reach the same conclusion. 

Now we show that it is impossible that ${\breve x_n} \in x^+$ and ${\hat x_n} \in {\bar x}^+$. Assume these relations hold and let $k$ be the least integer strictly greater than $n$ such that $x \in \Alph({{\hat x}_k})$ (such an integer exists since $\Delta_j$ is wavy). Let $\alpha_k = |{\hat x}_k|_x$ and $\beta_k = |{\hat x}_k|_{\bar x}$.
Since all of the words ${\hat x_n}$, \ldots, ${\hat x_{k-1}}$ belong to ${\bar x}^+$, we have
$\bar x {\bar v}_n {\hat x}_n {\bar v}_{n+1} \ldots {\bar v}_{k} {\hat x}_k 
\equiv \bar x {\bar v}_n {\hat x}_n {\bar v}_{n+1}\ldots {\bar v}_{k} x^{\alpha_k}{\bar x}^{\beta_k}
\equiv x v_n {\hat x}_n v_{n+1} \ldots v_k \bar x x^{\alpha_k-1}{\bar x}^{\beta_k}$.
Then by injectivity of the morphism $\mu_{x v_n}$, there exists an episturmian word directed by both  $\Delta = {\breve x_n}\prod_{m \geq n+1} v_m {\breve x_m}$ and $\Delta' = {\hat x}_n {v}_{n+1}\ldots {v}_{k} x^{\alpha_k-1}{\bar x}^{\beta_k}\bar x\prod_{m \geq k+1} {\bar v}_m {\hat x}_m$. But this is impossible since $\Delta$ directs a word starting with $x$ (recall that ${\breve x}_n \in x^+$) and $\Delta'$ directs a word starting with the first letter of $v_{n+1}$ (recall that ${\hat x}_n \in {\bar x}^+$).

\medskip

Let us now consider the case when one of the two words $\Delta_1$, $\Delta_2$ is wavy and the other has all of its spins ultimately $L$ or ultimately $R$. Then item~2 is verified by Proposition~\ref{P:JP_Prop3.6}.

\medskip
Suppose now that both $\Delta_1$ and $\Delta_2$ have all spins ultimately $L$ (resp.~ultimately $R$).
Then by Remark~\ref{R:spinnedVersions}, $\Delta_1$ and $\Delta_2$ are spinned versions of a common word. Hence $\Delta_1 = w\Delta$ and $\Delta_2 = w'\Delta$ (resp.~$\Delta_1 = w\bar\Delta$ and $\Delta_2 = w'\bar\Delta$) for some spinned words $w$, $w'$ of the same length and an infinite $L$-spinned word $\Delta$  (resp.~$R$-spinned word $\Delta$).  Since $\Delta_1$ and $\Delta_2$ are directive-equivalent, $\mu_w = \mu_{w'}$ by Proposition~\ref{P:JP_Prop3.7}, and furthermore $\Delta_1$ and $\Delta_2$ have infinitely many prefixes whose images are equal by $\mu$. Therefore, as already seen, this situation satisfies item~\ref{Ti:1}.

\medskip

We have now ended the study of the aperiodic case, since by Proposition~\ref{P:JP_Prop3.9}, $\Delta_1$ and $\Delta_2$ cannot direct a common aperiodic episturmian word if 
one of them has all spins ultimately $L$ and the other has all spins ultimately $R$. 

\medskip

Finally we come to the periodic case: 
$\Delta_1 = w \bx$ and $\Delta_2 = w'\by$ for some spinned words $w$, $w'$, letters $x$ and $y$,  and spinned infinite words $\bx \in \{x, \bar x\}^\omega$, $\by \in \{y, \bar y\}^\omega$. In this case, the episturmian word directed by $\Delta_1$ and $\Delta_2$ is $\mu_w(x)^\omega = \mu_{w'}(y)^\omega$, which implies that $\mu_w(x)^{|y|}= \mu_{w'}(y)^{|x|}$. Then (see \cite{mL83comb} for instance) there exists a primitive word $z$ such that $\mu_w(x)$ and $\mu_{w'}(y)$ are powers of $z$ (let us recall that a word $w$ is primitive if it is not an integer power of a shorter word, i.e., if $w = u^p$ with $p \in \NN$, then $p = 1$ and $w = u$). One can quite easily verify that any episturmian morphism maps any primitive word to another primitive word (see also \cite[Prop.~2.8, Prop.~3.15]{jJgP02epis}). Since any letter constitutes a primitive word, both $\mu_w(x)$ and $\mu_{w'}(y)$ are primitive. Thus $\mu_w(x) = z = \mu_{w'}(y)$. 
\end{proof}

\section{\label{S:normalization}Normalized directive word of an episturmian word}

In the previous section we have seen that any episturmian word $\bt$ has a directive word with infinitely many $L$-spinned letters.  
To work on Sturmian words, Berthé, Holton, and Zamboni recently proved that it is always possible to choose a particular directive word:

\begin{thrm}{\rm \cite{vBcHlZ06init}}
\label{theoDecSturm}
Any Sturmian word $\bw$ over $\{a,b\}$ has a unique representation of the form 
$$\displaystyle \bw = \lim_{n \rightarrow \infty}
L_a^{d_1-c_1}R_a^{c_1} L_b^{d_2-c_2}R_b^{c_2} \ldots
L_a^{d_{2n-1}-c_{2n-1}} R_a^{c_{2n-1}}
L_b^{d_{2n}-c_{2n}}R_b^{c_{2n}}(a)$$

where $d_k \geq
c_k \geq 0$ for all integer $k\geq 1$, $d_k \geq 1$ for $k \geq 2$ and
if $c_k = d_k$ then $c_{k-1} = 0$.
\end{thrm}

\medskip

In other words, any Sturmian word has a unique directive word over $\{a,b,\bar a, \bar b\}$ containing infinitely many $L$-spinned letters but no factor of the form ${\bar a}{\bar b}^na$ or ${\bar b}{\bar a}^nb$ with $n$ an integer. 
Actually this result is quite natural if one thinks about the presentation of the monoid of Sturmian morphisms (see \cite{See1991}). Using Theorems~\ref{T:presentation} and \ref{T:directSame}, we generalize Theorem~\ref{theoDecSturm} to episturmian words:

\begin{thrm}\label{T:normalisation}
Any episturmian word $\bt$ has a spinned directive word containing infinitely many $L$-spinned letters, but no factor in ${\bigcup}_{a\in\cA} \bar a \bar\cA^*a$. Such a directive word is unique if $\bt$ is aperiodic.
\end{thrm}

The example given in Remark~\ref{R:AboutDirectSame3} shows that unicity does not necessarily hold for periodic episturmian words. A directive word of an aperiodic episturmian word $\bt$ with the above property is called the {\em normalized directive word} of $\bt$. We extend this definition to morphisms: a finite spinned word $w$ is said to be a {\em normalized directive word} of the morphism $\mu_w$ if $w$ has no factor in ${\bigcup}_{a\in\cA} \bar a \bar\cA^*a$.

\medskip

One can observe that, by Theorem~\ref{T:presentation}, for any morphism in $L_a \mathcal{L}_\cA^\ast R_a$, we can find another decomposition of the morphism in the set $R_a \mathcal{R}_\cA^\ast L_a$. Equivalently, for any spinned word in $a\cA^*{\bar a}$, there exists a word $w'$ in ${\bar a}{\bar \cA}^*a$ such that $\mu_w = \mu_{w'}$. This is the main idea used in the proof of the lemma below. The proof of Theorem~\ref{T:normalisation} is based on an extension of this lemma to infinite words.

\medskip 

\begin{lmm}\label{L:normalisation_morph}
Any pure episturmian morphism has a unique normalized directive word.
\end{lmm}

\begin{proof}
{\em Existence of the normalized directive word}: 
Let $w = (w_i)_{1 \leq i \leq |w|}$ be a spinned word over $\cA \cup \bar \cA$.
We construct by induction on $|w|$ a normalized directive word of $\mu_{w}$. 

If $|w| = 0$, there is nothing to do: $\varepsilon$ is a normalized directive word of the empty morphism.
Assume we have constructed a normalized directive word $w' = (w'_i)_{1 \leq i \leq k}$ of the morphism $\mu_{w'} = \mu_{w}$.

Let $\bar x$ be a letter in ${\bar \cA}$. Then, by normalization of $w'$, the word $w' \bar x$ has no factor in $\cup_{a \in \cA} {\bar a} {\bar \cA}^\ast a$. Moreover since $\mu_{w} = \mu_{w'}$, we have $\mu_{w \bar x} = \mu_{w' \bar x}$: the word $w' \bar x$ is a normalized directive word of $\mu_{w \bar x}$.

Now let $x$ be a letter in $\cA$.
The word $w'x$ can have factors in $\cup_{a \in \cA} {\bar a} {\bar \cA}^\ast a$, but only as suffixes. If this does not hold, as in the previous case, the word $w'x$ is a normalized directive word of $\mu_{wx}$.
Else $w' = p {\bar x} {\bar u}_1 {\bar x} {\bar u}_2 \ldots {\bar x} {\bar u}_k$ for an integer $k \geq 1$, some $L$-spinned $x$-free words $({u}_i)_{1 \leq i \leq k}$ and a spinned word $p$ having no suffix in $\bar{x}\bar{\cA}^*$. 
The word $w''\bar{x}$ where $w'' = p x u_1 \bar{x} u_2 \ldots \bar{x} u_k$ contains no factor in $\cup_{a \in \cA} {\bar a} {\bar \cA}^\ast a$. Moreover Theorem~\ref{T:presentation} implies $\mu_{w'x} = \mu_{w''{\bar x}}$. Hence $w''{\bar x}$ is a normalized directive word of $\mu_{wx}$.

\medskip

Let us make a remark on the inductive construction presented in this proof: 
\begin{rmrk}
\label{R:onNormalization} Let $u, v, u', v'$ be four spinned words such that $u'$ (resp.~$v'$) is the normalized directive word obtained by the above construction from $u$ (resp.~$v$). If 
$u$ is a prefix of $v$ and if $p$ is a prefix of $u'$ ending by an $L$-spinned letter, then $p$ is also a prefix of $v'$.
\end{rmrk}

\medskip

{\em Unicity}: Assume by way of contradiction that $w$ and $w'$ are two different spinned normalized words such that $\mu_{w} = \mu_{w'}$. By left cancellativity of the monoid of episturmian morphisms, we can assume that $w$ and $w'$ start with different letters. Moreover it follows from Theorem~\ref{T:presentation} that $w$ and $w'$ are spinned versions of a common word. Without loss of generality, we can assume that $w$ begins with a letter $a \in \cA$ and $w'$ begins with $\bar{a}$ and so for any word $z$, $\mu_{w}(z) = \mu_{w'}(z)$ begins with $a$. Hence $w'$ must start with $\bar{a}\bar{v}a$ for a word $v \in \cA^*$. This contradicts its normalization.
\end{proof}

\medskip

\begin{xmpl}
Let $f$ be the pure episturmian morphism with directive word
${\bar a}{\bar b}c{\bar b}a{\bar b}{\bar a}{\bar c}{\bar b} {\bar a}{\bar c}a$.  
By Theorem~\ref{T:presentation},  
$\mu_{{\bar a}{\bar c}{\bar b}{\bar a}{\bar c}a} = 
 \mu_{{\bar a}{\bar c}{\bar b}ac{\bar a}} = 
 \mu_{acb{\bar a}c{\bar a}}$ and hence 
$\mu_{{\bar a}{\bar b}c{\bar b}a{\bar b}{\bar a}{\bar c}{\bar b} {\bar a}{\bar c}a} = 
\mu_{{\bar a}{\bar b}c{\bar b} a {\bar b}acb{\bar a}c{\bar a}}$
and ${\bar a}{\bar b}c{\bar b}a{\bar b}acb{\bar a}c{\bar a}$ is the normalized directive word of $f$.
\end{xmpl}

\medskip

Now we provide the

\medskip
\begin{proof2}{Theorem~\ref{T:normalisation}}

\noindent {\it Existence of the normalized directive word}:

Let $\Delta = (w_i)_{i \geq 1}$ be a spinned directive word of an episturmian word $\bt$ (with $w_i \in \cA\cup \bar\cA$). 
From Theorem~\ref{T:directSame}, we can assume that $\Delta$ has infinitely many $L$-spinned letters.

By Lemma~\ref{L:normalisation_morph}, for any $n \geq 1$, the morphism $\mu_{w_1 \ldots w_n}$ has a unique normalized directive word $(w^{(n)}_i)_{1 \leq i \leq n}$. (It follows from the proof of Lemma~\ref{L:normalisation_morph} that $w_i$ and $w^{(n)}_i$ are spinned versions of a common letter).

Let $p_n$ be the longest prefix of $w_1^{(n)} \ldots w_n^{(n)}$ that belongs to $(\cA \cup {\bar \cA})^\ast \cA$. Let $i_n \leq n$ be the integer such that $p_n= w_1^{(n)} \ldots w_{i_n}^{(n)}$, and let $\pi_n$ be the word $\pi_n=\mu_{w_1^{(n)} \ldots w_{i_n-1}^{(n)}}( w_{i_n}^{(n)})$.

Since the morphisms $\mu_{w_1 \ldots w_n}$ and $\mu_{w_1^{(n)} \ldots w_n^{(n)}}$ are equal, $\bt$ has the directive word $(w_1^{(n)}, \ldots, w_n^{(n)}, w_{n+1}, w_{n+2}, \ldots)$, so $\pi_n$ is a prefix of $\bt$.

By Remark~\ref{R:onNormalization}, for any $n \geq 1$, $p_n$ is a prefix of $p_{n+1}$, and since $\Delta$ contains infinitely many $L$-spinned letters, for any $n \geq 1$, there exists an $m > n$ such that $|p_m|>|p_n|$. 

If $|\Ult(\Delta)| = 1$, then there exists a letter $a$ and an integer $m$ such that $\bt = \mu_{p_m}(a^\omega)$ and $p_ma^\omega$ is a normalized directive word of $\bt$. If $|\Ult(\Delta)| > 1$, the sequence $(\pi_n)_{n \geq 1}$ is not ultimately constant, and $\lim_{n \rightarrow \infty} \pi_n=\bt$. In this case $\bt$ is directed by the sequence 
$\lim_{n \rightarrow \infty} p_n$ which is normalized by construction (indeed otherwise one of the prefixes $p_n$ would not be normalized). 

\medskip

\noindent {\it Unicity of the normalized directive word}:

Assume by way of contradiction that an aperiodic episturmian word $\bt$ has two different normalized spinned directive words $\Delta_1 = (w_n)_{n \geq 1}$ and $\Delta_2 = (w_n')_{n \geq 1}$ (with $w_n$ and  $w_n' \in \cA \cup {\bar \cA}$ for all $n$). Let $i \geq 1$ be the smallest integer such that $w_i \neq w_i'$ (and for all $j < i$, $w_j = w_j'$). By Theorem~\ref{T:directSame}, $\Delta_1$ and $\Delta_2$ are spinned versions of the  same word (see Remark~\ref{R:AboutDirectSame3}). Thus, without loss of generality, we can assume that $w_i=\bar{x}$ and $w_i'=x$ for some letter $x$. 

Let $\bt^{'(i)}$ be the episturmian word with (normalized) directive word $(w_n')_{n \geq i}$ (by Proposition~\ref{P:uniqueDirective} this word is unique), then $\bt^{'(i)}$ starts with $x$ since $w_i'=x$. Since the word $(w_n)_{n \geq 1}$ has infinitely many $L$-spinned letters, there exists an integer $j > i$ such that $w_j=y$ for a letter $y \in \cA$ and $w_{\ell} \in {\bar \cA}$ for each $\ell$, $i < \ell < j$. Let $\bt^{(i)}$ be the word with normalized directive word $(w_n)_{n \geq i}$, then $\bt^{(i)}$ has the word $\mu_{w_i \ldots w_{j-1}}(y)$ as prefix since $w_j=y$ and so $\bt^{(i)}$ starts with $y$ since $w_i \ldots w_{j-1} \in {\bar \cA}^\ast$. We have $\bt=\mu_{w_1 \ldots w_{i-1}}(\bt^{(i)})=\mu_{w_1' \ldots w_{i-1}'}(\bt^{'(i)})$. 
By choice of $i$, $w_1 \ldots w_{i-1}=w_1' \ldots w_{i-1}'$. Consequently, since episturmian morphisms are injective on infinite words, $\bt^{(i)} =\bt^{'(i)}$ and so $x=y$. But since $w_i=\bar{x}$, $w_{i+1} \ldots w_{j-1} \in {\bar \cA}^\ast$, and $w_j=x$, we reach a contradiction to the normalization of $(w_n)_{n \geq 1}$.
\end{proof2}

\section{\label{S:uniqueDirective}Episturmian words having a unique directive word}

In Section~\ref{S:equivalence} we have characterized pairs of words directing a common episturmian word. In Section~\ref{S:normalization} we have proposed a way to uniquely define any episturmian word through a normalization of its directives words (as mentioned in the introduction, see \cite{vBcHlZ06init,fLgR07quas,fLgR07quasB,aGfLgR07quas} for some uses of this normalization). Using these results we now characterize episturmian words having a unique directive word. 

\begin{thrm} \label{T:uniqueDirective}
An episturmian word has a unique directive word if and only if its (normalized) directive word contains 
1) infinitely many $L$-spinned letters,
2) infinitely many $R$-spinned letters, 
3) no factor in ${\bigcup}_{a\in\cA} \bar a \bar\cA^*a$, 
4) no factor in ${\bigcup}_{a\in\cA} a \cA^*\bar a$.

Such an episturmian word is necessarily aperiodic.
\end{thrm}

\begin{proof}
Assume first that an episturmian word $\bt$ has a unique spinned directive word $\Delta$. By Theorem~\ref{T:normalisation}, $\Delta$ is normalized and so contains infinitely many $L$-spinned letters and no factor in ${\bigcup}_{a\in\cA} \bar a \bar\cA^*a$. By item~3  of Theorem~\ref{T:directSame} and by Remark~\ref{R:periodic}, $\bt$ cannot be periodic. By item~\ref{Ti:2} of Theorem~\ref{T:directSame}, $\Delta$ also contains infinitely many $R$-spinned letters, and hence is wavy (otherwise one can construct another directive word of $\bt$ -- the fact that $\bt$ is aperiodic is important for having the $(v_n)_{n\geq 1}$ non-empty in this construction). Finally Theorem~\ref{T:presentation} implies the non-existence of a factor in ${\bigcup}_{a\in\cA} a \cA^*\bar a$ (otherwise, one can again construct another directive word for~$\bt$).

Let us now prove that the four conditions (given in the statement of the theorem) are sufficient. Arguing by contradiction, we assume that an episturmian word $\bt$ is directed by two spinned infinite words $\Delta_1$ and $\Delta_2$, both fulfilling the four given conditions. We observe that if $\Delta_1$ or $\Delta_2$ is ultimately written over $\{x, \bar{x}\}$ for a letter $x$ (which can occur only if $\bt$ is periodic), then at least one of the conditions is not fulfilled. Thus the two words $\Delta_1$ and $\Delta_2$ should verify one of the two first items in part $iii)$ of Theorem~\ref{T:directSame} (item 3 does not apply since $\bt$ is aperiodic).  But the hypotheses on $\Delta_1$ and $\Delta_2$ imply that only item~1 can be verified so that $\Delta_1 = \prod_{n \geq 1} v_n$, $\Delta_2 = \prod_{n \geq 1} z_n$ for spinned words $(v_n)_{n \geq 1}, (z_n)_{n \geq 1}$ such that $\mu_{v_n} = \mu_{z_n}$ for all $n \geq 1$. Now by Theorem~\ref{T:presentation} and by the fact that words $(v_n)_{n \geq 1}$ and $(z_n)_{n \geq 1}$ have  no factor in ${\bigcup}_{a\in\cA} \bar a \bar\cA^*a$ nor  ${\bigcup}_{a\in\cA} a \cA^*\bar a$, we must have $v_n = z_n$ for all $n \geq 1$. Thus $\Delta_1 = \Delta_2$.
\end{proof}

As an example, a particular family of episturmian words having unique directive words consists of those directed by {\em regular wavy words}, i.e., spinned infinite words having both infinitely many $L$-spinned letters and infinitely many $R$-spinned letters such that each letter occurs with the same spin everywhere in the directive word. More formally, a spinned version $\breve w$ of a finite or infinite word $w$ is said to be {\em regular} if, for each letter $x \in \Alph(w)$, all occurrences of $\breve x$ in $\breve w$ have the same spin $(L$ or $R)$. For example, $a\bar b a a \bar c \bar b$ and $(a\bar b c)^\omega$ are regular, whereas $a \bar b a \bar a \bar c b$ and $(a \bar b \bar a)^\omega$ are not regular.

\medskip
In the Sturmian case, we have:

\begin{prpstn} \label{P:uniqueDirective-Sturmian} 
Any Sturmian word has a unique spinned directive word or infinitely many spinned directive words. Moreover, a Sturmian word has a unique directive word if and only if its (normalized) directive word is regular wavy.
\end{prpstn}

\begin{proof} Let $\Delta$ be the normalized directive word of a Sturmian word $\bt$ over $\{a, b\}$. Then $\Delta$ contains no factor belonging to $\bar a \bar{b}^* a \cup  \bar b \bar{a}^* b$ (where $\alpha^* = \{\alpha\}^*$ for any letter $\alpha$). 

Assume first that $\Delta$ contains infinitely many factors in $ab^*\bar a \cup  b a^*\bar b$. Then $\Delta = p \bigcup_{n \geq 1} {x}_n {y}_n$  for some spinned words $p$ and $({x}_n$, ${y}_n)_{n \geq 1}$ such that, for all $n \geq 1$, $x_n \in ab^*\bar a \cup  ba^*\bar b$ and ${y}_n \in \{a, b, \bar a, \bar b\}^*$. In this case, $\Delta$ has infinitely many directive words; indeed, the spinned words $(p [\bigcup_{n = 1}^{k-1}{x}_n {y}_n] \bar{x}_k {y}_k \bigcup_{n \geq k+1} {x}_n {y}_n)_{_k \geq 1}$ are (by Theorem~ \ref{T:presentation}) pairwise different directive words for~$\bt$.

Now assume that $\Delta$ contains only finitely many factors in $ab^*\bar a \cup  b a^*\bar b$. Since $\Delta$ contains no factor in $\bar a \bar{b}^* a \cup  \bar b \bar{a}^* b$, it is ultimately regular wavy. More precisely $\Delta$ is regular wavy and either $\Delta$ belongs to $\{a, \bar b\}^\omega \cup \{\bar a, b\}^\omega$, or $\Delta$ belongs to one of the following sets of infinite words: $S_1 = \{a, b, \bar a, \bar b\}^* a \{\bar a, b\}^\omega$,
$S_2 = \{a, b, \bar a, \bar b\}^* b \{a, \bar b\}^\omega$,
$S_3 = \{a, b, \bar a, \bar b\}^* \bar a \{a, \bar b\}^\omega$ or
$S_4 = \{a, b, \bar a, \bar b\}^* \bar b \{\bar a, b\}^\omega$.
Assume $\Delta \in S_1$. Since any Sturmian word is aperiodic, $\Delta$ is not ultimately constant (see Remark~\ref{R:periodic}). Thus $\Delta = p a\bigcup_{n \geq 1} x_n \bar a$ with  $x_n \in {b}^*$ for all $n \geq 1$. Once again in this case, $\bt$ has infinitely many directive words since the words 
$({p}[\bigcup_{n = 1}^{k-1} \bar a\bar{x}_n]a\bigcup_{n \geq k} x_n \bar a)_{k \geq 1}$ are pairwise different directive words for $\bt$. The cases when $\Delta \in S_2$ or $\Delta \in S_3$ or $\Delta \in S_4$ are similar.

We end with the case when $\Delta$ is regular wavy. In this case, $\Delta$ contains infinitely many $L$-spinned letters, infinitely many $R$-spinned letters, no factor in $ab^*\bar{a} \cup ba^*\bar{b}$, and no factor in 
$\bar a \bar{b}^* a \cup  \bar b \bar{a}^* b$. Hence by Theorem~ \ref{T:uniqueDirective}, $\bt$ has a unique directive word.
\end{proof}

Proposition~\ref{P:uniqueDirective-Sturmian} shows a great difference between Sturmian words and episturmian words constructed over alphabets with at least three letters. Indeed, when considering words over a ternary alphabet, one can find episturmian words having exactly $m$ directive words  for any $m \geq 1$. For instance, the episturmian word $\bt$ directed by $\Delta = a(b\bar a)^{m-1}b\bar c(ab\bar c)^\omega$ has exactly $m$ directive words, namely $(\bar a\bar b)^ia(b\bar a)^{j}b\bar c(ab\bar c)^\omega$ with $i+j = m-1$. Notice that the suffix $b\bar c(ab\bar c)^\omega$ of $\Delta$ is regular wavy, and the other $m-1$ spinned versions of $\Delta$ that also direct $\bt$ arise from the $m-1$ words that are block-equivalent to the prefix $a(b\bar a)^{m-1}$. 

\medskip

\noindent
\textbf{Acknowledgement}.
The two last authors thank Eddy Godelle for his remarks and his suggestion for improvement of the proof of Theorem~\ref{T:normalisation}.

\footnotesize
\bibliographystyle{elsart-num-sort}
\bibliography{GLR}

\begin{thebibliography}{10}
\expandafter\ifx\csname url\endcsname\relax
  \def\url#1{\texttt{#1}}\fi
\expandafter\ifx\csname urlprefix\endcsname\relax\def\urlprefix{URL }\fi

\bibitem{AS2003}
J.-P. Allouche, J.~Shallit, Automatic sequences: Theory, Applications,
  Generalizations, Cambridge University Press, 2003.

\bibitem{pAgR91repr}
P.~Arnoux, G.~Rauzy, Repr\'esentation g\'eom\'etrique de suites de
  complexit\'es $2n+1$, Bull. Soc. Math. France 119 (1991) 199--215.

\bibitem{jB07stur}
J.~Berstel, Sturmian and {e}pisturmian {w}ords (a survey of some recent
  results), in: Proceedings of CAI 2007, vol. 4728 of Lecture Notes in Computer
  Science, Springer-Verlag, 2007.

\bibitem{vBcHlZ06init}
V.~Berth\'e, C.~Holton, L.~Q. Zamboni, Initial powers of {S}turmian sequences,
  Acta Arith. 122 (2006) 315--347.

\bibitem{xDjJgP01epis}
X.~Droubay, J.~Justin, G.~Pirillo, Episturmian words and some constructions of
  de {L}uca and {R}auzy, Theoret. Comput. Sci. 255~(1-2) (2001) 539--553.

\bibitem{sF99comp}
S.~Ferenczi, Complexity of sequences and dynamical systems, Discrete Math. 206
  (1999) 145--154.

\bibitem{aG06onst}
A.~Glen, On {S}turmian and episturmian words, and related topics, Ph.D. thesis,
  The University of Adelaide, Australia (April 2006).

\bibitem{aG06acha}
A.~Glen, A characterization of fine words over a finite alphabet, Theoret.
  Comput. Sci. 391 (2008) 51--60.

\bibitem{aGjJ07epis}
A.~Glen, J.~Justin, Episturmian words: a survey, preprint, 2007.

\bibitem{aGjJgP06char}
A.~Glen, J.~Justin, G.~Pirillo, Characterizations of finite and infinite
  episturmian words via lexicographic orderings, European J. Combin. 29 (2008)
  45--58.

\bibitem{aGfLgR07quas}
A.~Glen, F.~Lev\'e, G.~Richomme, Quasiperiodic and {L}yndon episturmian words,
  preprint, 2008.

\bibitem{eG06auto}
E.~Godelle, Repr\'esentation par des transvections des groupes d'artin-tits,
  Group, Geometry and Dynamics 1 (2007) 111--133.

\bibitem{jJgP02epis}
J.~Justin, G.~Pirillo, Episturmian words and episturmian morphisms, Theoret.
  Comput. Sci. 276~(1-2) (2002) 281--313.

\bibitem{jJgP02onac}
J.~Justin, G.~Pirillo, On a characteristic property of {A}rnoux-{R}auzy
  sequences, Theoret. Inform. Appl. 36~(4) (2003) 385--388.

\bibitem{jJgP04epis}
J.~Justin, G.~Pirillo, Episturmian words: shifts, morphisms and numeration
  systems, Internat. J. Found. Comput. Sci. 15~(2) (2004) 329--348.

\bibitem{fLgR04quas}
F.~Lev\'e, G.~Richomme, Quasiperiodic infinite words: some answers, Bull. Eur.
  Assoc. Theor. Comput. Sci. 84 (2004) 128--138.

\bibitem{fLgR07quasB}
F.~Lev\'e, G.~Richomme, Quasiperiodic episturmian words, in: Proceedings of the
  $6$th International Conference on Words, Marseille, France, September 17-21,
  2007.

\bibitem{fLgR07quas}
F.~Lev\'e, G.~Richomme, Quasiperiodic {S}turmian words and morphisms, Theoret.
  Comput. Sci. 372 (2007) 15--25.

\bibitem{mL83comb}
M.~Lothaire, Combinatorics on Words, vol.~17 of Encyclopedia of Mathematics and
  its Applications, Addison-Wesley, 1983.

\bibitem{mL02alge}
M.~Lothaire, Algebraic Combinatorics on Words, vol.~90 of Encyclopedia of
  Mathematics and its Applications, Cambridge University Press, 2002.

\bibitem{MH1940}
M.~Morse, G.~Hedlund, Symbolic {D}ynamics {II}. {S}turmian trajectories, Amer.
  J. Math. 61 (1940) 1--42.

\bibitem{PV2006}
G.~Paquin, L.~Vuillon, A characterization of balanced episturmian sequences,
  Electron. J. Combin. 14, \#R33, pp.~12.

\bibitem{nP02subs}
N.~Pytheas~Fogg, Substitutions in Dynamics, Arithmetics and Combinatorics, vol.
  1794 of Lecture Notes in Mathematics, Springer, 2002.

\bibitem{gR82nomb}
G.~Rauzy, Nombres alg\'ebriques et substitutions, Bull. Soc. Math. France 110
  (1982) 147--178.

\bibitem{gR85mots}
G.~Rauzy, Mots infinis en arithm\'etique, in: M.~Nivat, D.~Perrin (eds.),
  Automata on Infinite words, vol. 192 of Lecture Notes in Computer Science,
  Springer-Verlag, Berlin, 1985.

\bibitem{gR03conj}
G.~Richomme, Conjugacy and episturmian morphisms, Theoret. Comput. Sci.
  302~(1-3) (2003) 1--34.

\bibitem{gR03lynd}
G.~Richomme, Lyndon morphisms, {B}ull. {B}elg. {M}ath. {S}oc. Simon Stevin 10
  (2003) 761--785.

\bibitem{gR07conj}
G.~Richomme, Conjugacy of morphisms and {L}yndon decomposition of standard
  {S}turmian words, Theoret. Comput. Sci. 380~(3) (2007) 393--400.

\bibitem{gR07aloc}
G.~Richomme, A local balance property of episturmian words, in: Proc. DLT '07,
  vol. 4588 of Lecture Notes in Computer Science, Springer, Berlin, 2007, pp.
  371--381.

\bibitem{rRlZ00agen}
R.~Risley, L.~Zamboni, A generalization of {Sturmian} sequences:
  {combinatorial} structure and transcendence, Acta Arith. 95 (2000) 167--184.

\bibitem{See1991}
P.~S\'e\'ebold, Fibonacci morphisms and {S}turmian words, Theoret. Comput. Sci.
  88~(2) (1991) 365--384.

\bibitem{zWyZ99some}
Z.-X. Wen, Y.~Zhang, Some remarks on invertible substitutions on three letter
  alphabet, Chinese Sci. Bull. 44~(19) (1999) 1755--1760.

\end{thebibliography}
 
\end{document}